\documentclass[superscriptaddress,prl,twocolumn,
amsmath,amssymb,showpacs]{revtex4}
\usepackage{graphicx}

\begin{document}

\title{
Enhanced Pairing Correlations near Oxygen Dopants 
in Cuprate Superconductors} 

\author{Giniyat Khaliullin}
\affiliation{Max-Planck-Institut f\"ur Festk\"orperforschung,
Heisenbergstrasse 1, D-70569 Stuttgart, Germany}

\author{Michiyasu Mori}
\affiliation{Advanced Science Research Center, Japan Atomic Energy Agency, 
Tokai-mura, Ibaraki 319-1195, Japan}
\affiliation{CREST, Japan Science and Technology Agency,
Sanbancho 102-0075, Japan}

\author{Takami Tohyama}
\affiliation{Yukawa Institute for Theoretical Physics, Kyoto University,
Kyoto 606-8502, Japan}

\author{Sadamichi Maekawa}
\affiliation{Advanced Science Research Center, Japan Atomic Energy Agency, 
Tokai-mura, Ibaraki 319-1195, Japan}
\affiliation{CREST, Japan Science and Technology Agency,
Sanbancho 102-0075, Japan}

\begin{abstract}
Recent experiments on Bi-based cuprate superconductors have revealed an 
unexpected enhancement of the pairing correlations near the interstitial 
oxygen dopant ions. Here we propose a possible mechanism -- based on local
screening effects -- by which the oxygen dopants do modify the electronic 
parameters within the CuO$_2$ planes and strongly increase the superexchange 
coupling $J$. This enhances the spin pairing effects locally and may explain 
the observed spatial variations of the density of states and the pairing gap. 
\end{abstract}

\date{\today}


\maketitle

In terms of global phase behavior, the cuprate families hosting high-$T_c$ 
superconductivity (SC) are similar: parent compounds are insulating 
antiferromagnets with $T_N\sim 300-400K$, and become SC once doped 
charge carriers destroy magnetic order and fermionic bands are developed. 
This universality manifests that the interactions driving 
magnetism and superconductivity are mostly confined to the CuO$_2$ planes
common to all the high-$T_c$ cuprates. What is remarkable, however, is that the 
SC transition temperature $T_c$ broadly varies from one family to another, 
ranging from $\sim 40K$ in La$_{2-x}$Sr$_x$CuO$_4$ to the values
as high as $\sim 140K$ in Hg-based compounds. A number of ideas concerning the 
material dependence of $T_c$ have been discussed: the influence of apical 
oxygens \cite{Oht91}, charge ordering effects \cite{Bas00}, band structure 
variations \cite{Pav01}, the type of dopant induced disorder \cite{Eis04}, 
etc. Given the high sensitivity of correlated electrons in oxides to external 
perturbations, all these material specific factors may indeed have a strong 
impact on how the competition between SC and magnetism is resolved in a 
particular compound and thus on the $T_c$ values.  

Local disorder -- an inevitable side effect of chemical doping -- 
triggers also nanoscale phase inhomogeneities commonly observed in cuprates.  
Signatures of the pairing effects well above the bulk $T_c$ (see, e.g., 
\cite{Li10,Gom07,Pas08}) have led to a notion of ''local pairing gaps'' 
and ''local $T_c$''. Typically, the dopant ions are expected to perturb the 
lattice and chemical bonds, reduce the electronic mobility, and thus suppress 
SC in favor of magnetism \cite{Eis04}. Strong enhancement of the pairing 
correlations near the interstitial oxygen dopant ions in 
Bi$_2$Sr$_2$CaCu$_2$O$_{8+\delta}$ reported by McElroy {\it et al.}
\cite{McE05} was therefore a big surprise. Their scanning tunneling 
microscopy (STM) data have been nicely reproduced by the model calculations 
\cite{Nun05} assuming a positive impact of the dopant ions on the pairing 
potential, and triggered a broad discussion 
\cite{He06,Mas07,Pet09,Foy09,Joh09,Oka10} 
on the origin of this striking observation. 

Given the essential role of magnetic interactions in cuprates, there has been 
a strong focus on the possible increase of the spin exchange coupling $J$ by
dopant ions. In principle, this ''quantum-chemistry'' parameter is sensitive 
to local electronic structure and widely varies in cuprates, from 
$\sim 110-140$~meV in SC cuprates to the values as high as $\sim 240$~meV 
in a quasi-one dimensional compound Sr$_2$CuO$_3$ (see Ref.~\cite{Mae04} 
for related discussions). However, the most detailed calculations 
\cite{Joh09,Foy09} found so far only moderate impact of dopants on $J$. 

In this Letter, we propose a mechanism by which the dopant oxygens may indeed 
strongly enhance the interaction $J$ locally. Our key idea is to go beyond
static level structure used in standard calculations of $J$, by taking
into account a dynamical change of electronic parameters due to polarization 
effects. Different from previous work considering a dopant 
ion as mere point charge modifying the Madelung potentials on CuO$_2$ planes 
(related changes in $J$ are small \cite{Joh09,Foy09}), we notice that there 
is a strong covalency between the dopant O$_d$ and closely located apical 
O$_a$ oxygen electrons, forming a molecular orbital complex (see Fig.~1). 
In fact, the mixing of O$_d$ and apical O$_a$ orbitals is suggested by 
experiment \cite{Ric06}. Consider now virtual $p$-$d$ and $d$-$d$ charge 
transitions within the Cu-O-Cu bond that lead to the spin exchange $J$. 
We will show that the corresponding excitation energies $\Delta_{pd}$ and 
$U$ are dynamically screened by the polarization of molecular orbitals hence 
enhancing $J$. The ionic polarization effects on excitation energies  
are known \cite{Boe84,Bri95,Mor08}; here, the effect is greatly amplified 
due to cooperative response of the spatially extended O$_a-$O$_d-$O$_a$ complex 
having much higher polarizability than that of constituent ions alone. 
We will also show, by an exact diagonalization of the $t-J$ model, that 
local enhancement of $J$ leads to the spatial variations in density of 
electronic states (DOS) observed in STM experiments \cite{Pas08}. Our 
findings suggest an interesting possibility of quantum-chemistry control 
of the key interaction $J$ in cuprates.     

To begin with, we recall that a dopant oxygen O$_d$ is located between 
the SrO and BiO planes \cite{McE05,He06}, at the distance $\sim$~2.2~\AA~from 
two apical O$_a$ ions. In the ground state, i.e., before the virtual charge 
excitation within the CuO$_2$ plane is made, Hamiltonian of the 
O$_a-$O$_d-$O$_a$ molecule comprises two terms, 
$H_{mol} = H_{ion} + H_{cov}$, where the first term  
\begin{equation}
H_{ion}=[E^{(d)}_p n_p + E^{(d)}_s n_s] + \sum_{a=1,2}
[E_p n^{(a)}_p + E_s n^{(a)}_s] 
\label{Hion}
\end{equation}
stands for the $2p$ and $3s$ electron energies of a dopant ($E^{(d)}_{p,s}$) and 
apical ($E_{p,s}$) oxygen ions: $n_p=p^{\dag}p$ and $n_s=s^{\dag}s$ are the 
corresponding particle numbers. Experimental data \cite{McE05,Ric06} and 
band structure calculations \cite{He06} suggest that $2p$ level 
of a dopant O$_d$ is higher than that of apical O$_a$ by several eV's; 
we will take below $E^{(d)}_p-E_p =2$~eV (the results are not very sensitive 
to this parameter). In Eq.~(\ref{Hion}), summation over the spin direction 
and $2p$ orbitals lying in the $xy$ plane formed by the O$_a-$O$_d-$O$_a$ 
molecule, see Fig.~1(a), is implied. 

\begin{figure}[tbp]
\includegraphics[width=7.5cm]{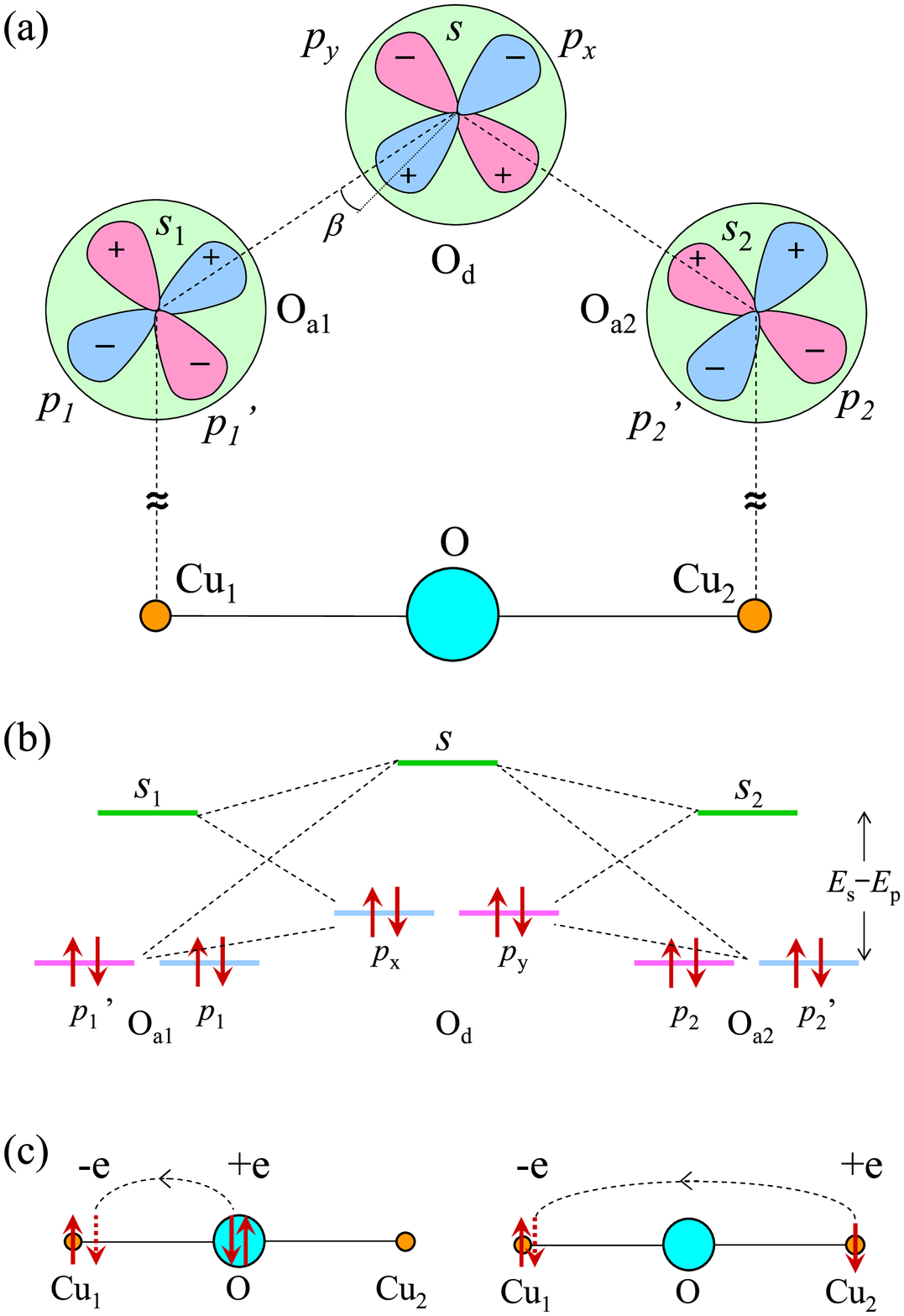}
\caption{(color online). 
(a)~The O$_a-$O$_d-$O$_a$ molecular complex formed by a dopant O$_d$ and its 
two apical O$_a$ oxygen neighbors. $2p$ and $3s$ orbitals involved in the 
polarization process are depicted. 
(b)~The level structure and hopping pathways of these orbitals. 
(c)~Charge disproportionation due to the $p$-$d$ (left) and $d$-$d$ (right) 
electron transfer within the CuO$_2$ plane. The corresponding excitation 
energies $\Delta_{pd}$ and $U$ are screened by the high-energy transitions 
between O$_a-$O$_d-$O$_a$ molecular orbitals.
}
\label{fig1}
\end{figure}

The second term in $H_{mol}$ describes the formation of the molecular orbitals 
due to the hopping between the dopant and apical oxygen states: 
\begin{eqnarray}
H_{cov}=
&-&T_{pp\sigma} [p^\dag_x (up_1-vp_2) + p^\dag_y (up_2-vp_1)] \label{Hcov}\\ 
&-&T_{pp\pi} [p^\dag_x (up'_2+vp'_1) + p^\dag_y (up'_1+vp'_2)] \nonumber\\
&-&T_{sp} [p^\dag_x (us_1-vs_2) + p^\dag_y (us_2-vs_1)] \nonumber\\
&-&T_{sp} [s^\dag (p_1+p_2)] - T_{ss} [s^\dag (s_1+s_2)] + h.c. , \nonumber 
\end{eqnarray}
where $u=\cos\beta$, $v=\sin\beta$, and $\beta$ is the angle between 
the apical $p_1$ and dopant $p_x $ orbitals [see Fig.~1(a)]. 
Strong mixing of electronic states via $H_{cov}$ enhances the polarizability of 
the O$_a-$O$_d-$O$_a$ molecular complex, the effect which goes well beyond a 
conventional dipolar polarization of oxygen ions alone \cite{Boe84,Bri95,Mor08}.

We use the relation $T_{pp\pi}=-\frac{1}{2}T_{pp\sigma}$ \cite{And78} and 
denote $T_{pp\sigma} \equiv T_{pp}$. The signs in Eq.~(\ref{Hcov}) are 
dictated by the orbital structure in Fig.~1(a), with a convention 
$T_{pp}, T_{sp}, T_{ss}>0$. Since 3$s$ orbital is more extended than 
2$p$ one, a relation $T_{pp} < T_{sp} < T_{ss}$ holds \cite{And78}. We assume 
the scaling $(T_{pp}, T_{sp}, T_{ss})= T_{pp}(1, \kappa, \kappa^2)$ with a 
representative value $\kappa=4/3$, and vary $T_{pp}$ within the 
$1.5-2.0$~eV range. This should be a reasonable estimate for the 
overlap between rather extended 2$p$ oxygen orbitals at distance 
$R(O_a-O_d)\simeq$~2.2~\AA; for comparison, the hopping $t_{pd\sigma}$ 
between the less extended copper $d_{3x^2-r^2}$ and in-plane oxygen 2$p_x$ 
orbitals at similar distance $R(Cu-O)\simeq$~1.9~\AA~is about 
$-1.8$~eV \cite{McM88}. 
 
The exchange $J$ between Cu spins is realized via the virtual hoppings 
of electrons in the CuO$_2$ planes. The charges dynamically generated 
on Cu and O sites during these transitions [see Fig.~1(c)] modify 
the energy levels of $2p$ and $3s$ electrons on the apical and dopant 
oxygens: $E^{(a)}_{p,s}(\varphi)=E_{p,s} + \varphi(\vec{r}_a)$ and 
$E^{(d)}_{p,s}(\varphi)=E^{(d)}_{p,s} + \varphi(\vec{r}_d)$, correspondingly. 
The energy shifts $\varphi(\vec{r})$ are determined by Coulomb potentials of 
the virtual charges. For instance, the $p$-$d$ hopping process creates 
an extra electron (hole) on Cu (O) sites [Fig.~1(c) left]. The potentials 
$\varphi(\vec{r}_a), \varphi(\vec{r}_d)$ of these charges cause a dynamical 
reconstruction of the molecular orbitals and their populations, i.e., 
polarize the O$_a-$O$_d-$O$_a$ complex, which results in the renormalization 
of the virtual excitation energies. 
  
Under the 
potential $\varphi(\vec{r})$ of virtual charges, 
the Hamiltonian of the O$_a-$O$_d-$O$_a$ complex becomes 
$H_{mol}(\varphi) = H_{ion}(\varphi) + H_{cov}$, where 
$H_{ion}(\varphi) = H_{ion} + H_{\varphi}$ with 
\begin{eqnarray}
H_{\varphi} 
&=& \varphi(\vec{r}_d)(n_p+n_s)+
\sum_{a=1,2} \varphi(\vec{r}_a)(n^{(a)}_p + n^{(a)}_s) 
\nonumber\\
&+& \mu \sum_{a=1,2} \nabla \varphi(\vec{r}_a)
[s_a^\dag (p_a \vec{i}+p'_a\vec{j}) + h.c.].
\label{Hphi}
\end{eqnarray}
The first two terms stand for the energy shifts of 2$p$ and 3$s$ levels, while 
the gradient term describes the ionic polarization of apical oxygens 
(which are close to the CuO$_2$ plane). Here, $\mu=\langle 2p|r|3s \rangle$ 
is the dipolar matrix element and the unit vectors $\vec{i}$ and $\vec{j}$ 
select a proper 2$p$ orbital oriented along the gradient $\nabla\varphi$. 
The value of $\mu$ can be estimated from the polarizability $\alpha$ of 
O$^{2-}$ ion \cite{Boe84}: 
$\mu\simeq\frac{1}{2e}\sqrt{\alpha(E_s-E_p)}\simeq 0.5-0.6$~\AA, 
considering $\alpha\simeq$~2~\AA$^3$ \cite{Sha93} and the 3$s$--2$p$ level 
separation $E_s-E_p\simeq 7 - 10$~eV \cite{noteE}. The gradient term has 
been included because it may change the orbital populations; however, this
purely ionic effect is found to be small. 

We assume that the electronic transitions within the O$_a-$O$_d-$O$_a$ complex
are faster than charge fluctuations in Cu--O--Cu exchange bonds. Indeed, 
the (inverse) time scale for the polarization process is determined by 
3$s$-2$p$ level separation $\sim10$~eV. This is much larger than the Cu--O 
charge-transfer energy $\Delta_{pd}\sim 3$~eV which is the dominant parameter 
dictating the strength of Cu--Cu virtual hopping $t=t_{pd}^2/\Delta_{pd}$ 
\cite{notet} and thus $J$. Approximation of an ''instantaneous'' polarization 
simplifies the calculations enabling us to express the screening effects in 
terms of the energies of O$_a-$O$_d-$O$_a$ complex in the initial 
[$\varphi(\vec{r})=0$] and intermediate [$\varphi(\vec{r})\neq 0$] states, 
i.e., before and after the electron hopping in the CuO$_2$ plane is made. For a 
given transition, e.g., Cu-O $p$-$d$ one, we consider the corresponding 
Coulomb potentials $\varphi(\vec{r})$ on O$_{1,2}$ and O$_d$ ions and 
calculate the energies $E_{mol}^{pol}=\langle H_{mol}(\varphi)-H_{mol}\rangle$ 
and $E_{ion}^{pol}=\langle H_{ion}(\varphi)-H_{ion}\rangle$. 
These quantities represent the polarization corrections to the $p$-$d$ charge- 
transfer energy, with the O$_a-$O$_d-$O$_a$ covalency effects included or 
only considering the ionic polarization, correspondingly. The difference 
$\delta E = E_{mol}^{pol} - E_{ion}^{pol}$ is precisely what we are looking
for: the renormalization of $\Delta_{pd}$ due to high-energy response of the 
O$_a-$O$_d-$O$_a$ complex. We find $E_{mol}^{pol} \gg E_{ion}^{pol}$; i.e., 
the effect is indeed dominated by the polarization of molecular orbitals.  

The coupling $J$ is estimated from a familiar expression:  
\begin{equation}
J = \frac{4t_{pd}^4}{\Delta_{pd}^2} 
\left(\frac{1}{U}+\frac{1}{\Delta_{pd}+\frac{1}{2}U_p}\right).    
\label{J}
\end{equation}
Three distinct intermediate states, generated by $p$-$d$ 
($\Delta_{pd}$ process), $d$-$d$ ($U$ process), and so-called pair hopping 
($2\Delta_{pd}$ process with two holes on oxygen) transitions, are involved 
here. For each state, we use the corresponding potential $\varphi(\vec{r})$ 
and evaluate the polarization corrections to the excitation energies.  

\begin{figure}[tbp]
\includegraphics[width=8.5cm]{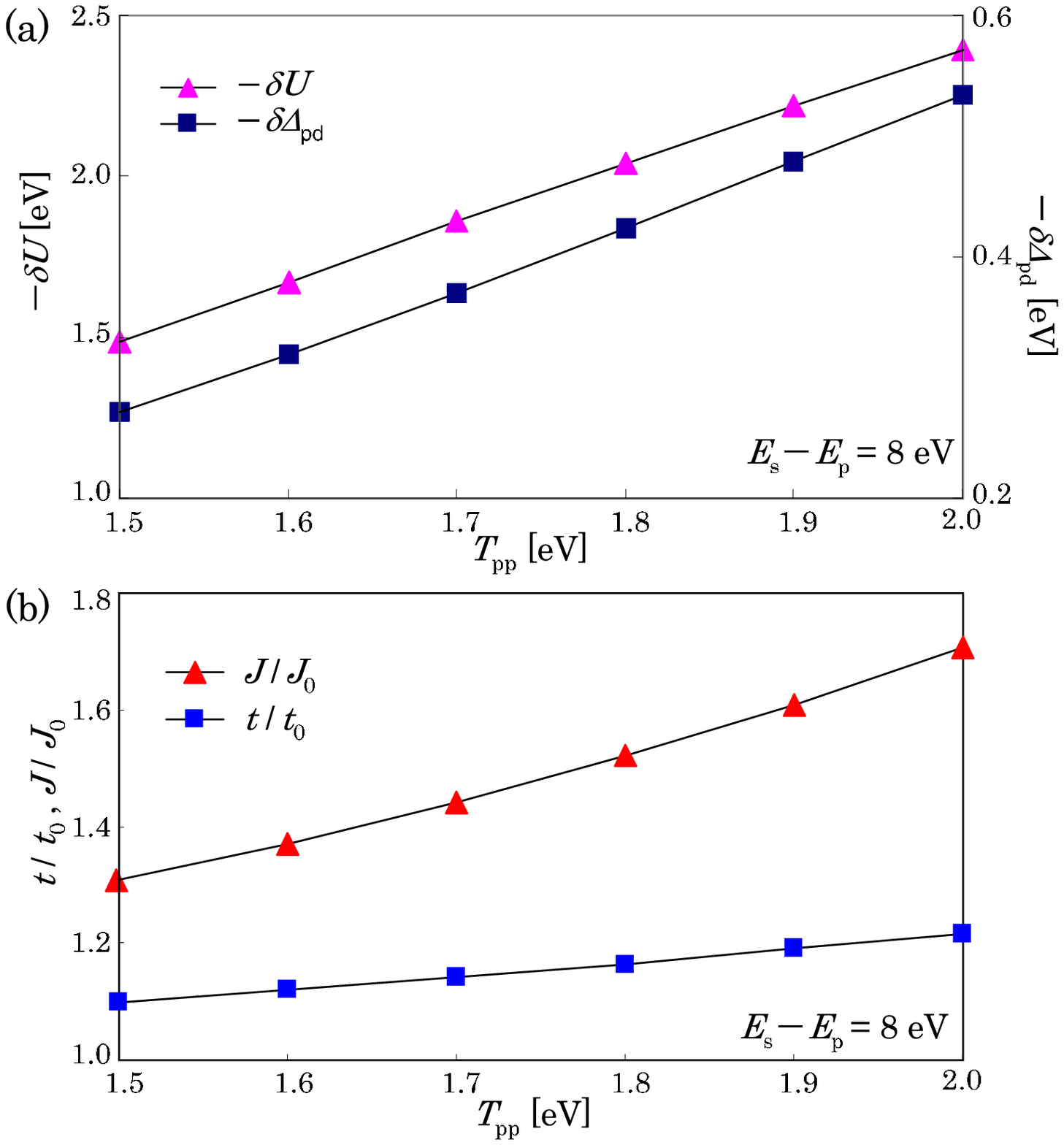}
\label{fig2}
\caption{(color online). 
(a)~Reduction of the $\Delta_{pd}$ and $U$ excitation energies due to a
polarization of the O$_a-$O$_d-$O$_a$ molecular complex, as a function of 
hopping $T_{pp}$. 
(b)~Enhancement of a virtual hopping $t=t_{pd}^2/\Delta_{pd}$ and the 
superexchange coupling $J$ on the Cu-Cu bond below a dopant oxygen. 
}
\end{figure}

As shown in Figs.~2(a) and 3(a), the dopant has a surprisingly strong 
impact on $\Delta_{pd}$ and $U$ values (however, the correction to the 
$2\Delta_{pd}-$process is negligible). The effect is sensitive to the 
hopping $T$ between the apical and dopant orbitals and to the energy 
$E_s-E_p$ between an empty and occupied states: both parameters directly 
control the strength of covalency within the O$_a-$O$_d-$O$_a$ molecule 
and thus its polarizability. It is interesting to note that doping by the 
larger size ions like sulfur S should increase the covalency and 
enhance the effect.     

\begin{figure}[tbp]
\includegraphics[width=8.5cm]{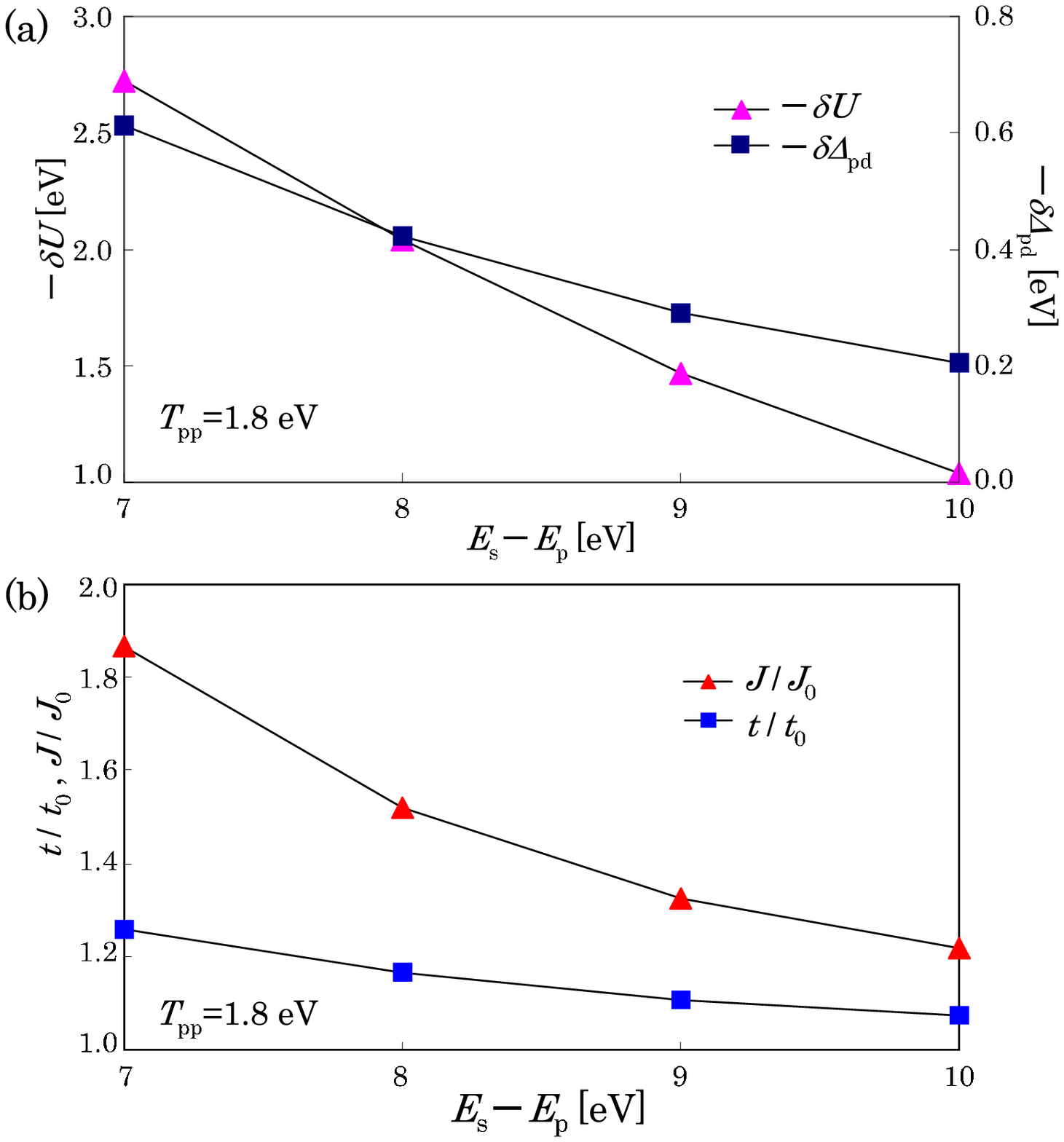}
\caption{(color online). 
(a)~Reduction of the $\Delta_{pd}$ and $U$ excitation energies 
as a function of 3$s$-2$p$ energy level separation $E_s-E_p$.
(b)~Enhancement of $t$ and $J$ values on the Cu-Cu bond below a dopant ion. 
}
\label{fig3}
\end{figure}

Having obtained the polarization corrections to excitation energies, we may 
address the $J$ value, Eq.~(\ref{J}). We assume that $p$-$d$ hopping $t_{pd}$ 
is not affected by dopants \cite{note_tpd}, and calculate the ratio $J/J_0$ of 
the spin coupling $J$ on the bond below a dopant oxygen to bare $J_0$ in the 
``bulk'', using $\Delta_{pd}=3$~eV, $U=8$~eV, and $U_p=4$~eV as the unscreened 
values. The results are presented in Figs.~2(b) and 3(b), together with 
$t/t_0=\Delta_{pd}/(\Delta_{pd}+\delta\Delta_{pd})$ values (note that 
$\delta\Delta_{pd}<0$). A dramatic, by a factor of 1.5 or even larger, 
enhancement of the interaction $J$ is observed, mainly due to an increased 
effective hopping $t\propto1/\Delta_{pd}$ (note that $J\propto t^2$).  

The above results demonstrate a pronounced effect of the oxygen
dopants on magnetic interactions which are believed to be relevant for 
the pairing in cuprates. To make a closer link with the experiment, we have
calculated a local DOS by an exact diagonalization of the $t-J$ model 
(on 20-site cluster with 2 holes) where a particular bond has an 
enhanced $J$ value \cite{noteJ}. Reduction of the DOS at the Fermi level 
and strong shift of the hump feature to higher binding energy is found for 
sites with large $J$ coupling (see Fig.~4). This is consistent with  
the normal state data by Pasupathy {\it et al.} \cite{Pas08}, and 
has a clear interpretation in terms of an enhanced spin singlet pairing near 
the dopants. Based on the model calculations by Nunner {\it et al.} 
\cite{Nun05}, it is natural to think that the same correlations are also 
responsible for the local enhancement of the pairing gaps in the SC 
state \cite{McE05,Pas08}; however, this remains to be clarified since our 
small cluster calculations cannot directly address the SC state properties.

\begin{figure}[tbp]
\includegraphics[width=8.4cm]{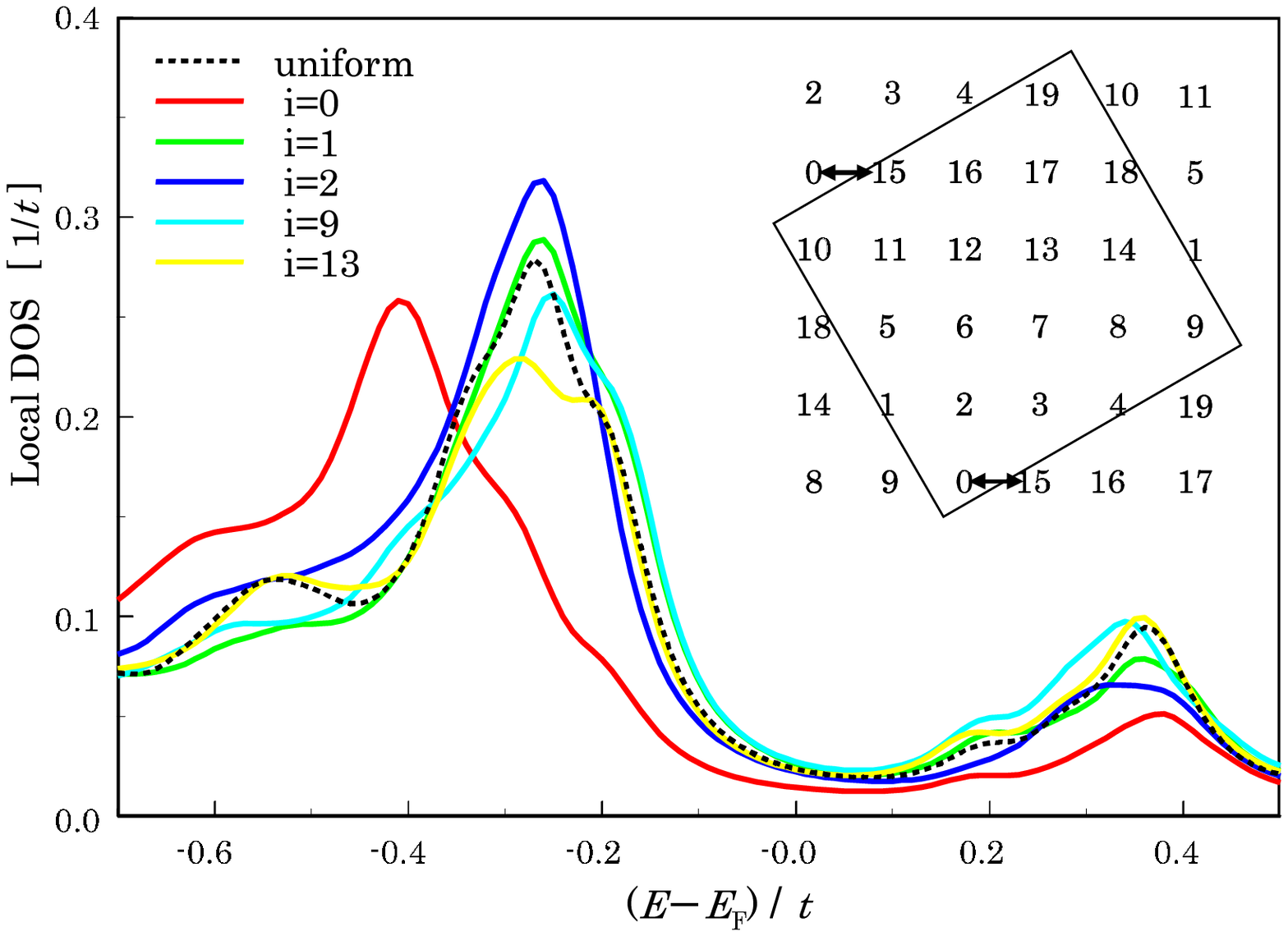}
\caption{(color online). 
Exact diagonalization results for the local DOS (per spin direction) 
on different sites $i$ (=0,1,...) of the $t-t'-J$ model cluster 
shown in the inset. The periodic boundary conditions are applied. 
The next-nearest-neighbor hopping $t'=-0.4t$. One particular exchange bond 
indicated by an arrow has an enhanced $J=1.7J_0$ (compared to $J_0=0.4t$ in 
the ''bulk''). On the corresponding sites $i=0$ or $i=15$, the spectral 
weight is suppressed at the Fermi level and transferred to the higher binding 
energies. The result for $J=J_0$ case is represented by a dotted line.
}
\label{fig4}
\end{figure}

To summarize, the oxygen dopants have a strong positive impact on magnetic 
interactions $J$ and enhance the pairing effects locally. The effect is of 
an entirely dynamical nature and thus goes beyond the band structure 
calculations. Apart from the STM experiments, there is an interesting 
possibility to test our theory by the resonant inelastic x-ray scattering: 
it monitors the high-energy spin excitations \cite{Bra10}, and a broad 
distribution of $J$ values predicted here can be directly observed.  

We thank P. Hirschfeld, T. Hanaguri, A. Yazdani, and B. Keimer 
for discussions. This work was supported by Grant-in-Aid for 
Scientific Research and Next-Generation Supercomputer Project from MEXT. 
G.Kh. thanks YIPQS of YITP at Kyoto University, IMR at Tohoku University, 
and ASRC at JAEA for kind hospitality. 



\begin{thebibliography}{99}

\bibitem{Oht91}
Y. Ohta, T. Tohyama, and S. Maekawa,
Phys. Rev. B {\bf 43}, 2968 (1991); R. Raimondi, J.H. Jefferson, and
L.F. Feiner, Phys. Rev. B {\bf 53}, 8774 (1996). 

\bibitem{Bas00} 
G. Baskaran, Mod. Phys. Lett. B {\bf 14}, 377 (2000).

\bibitem{Pav01}
E. Pavarini {\em et al.},
Phys. Rev. Lett. {\bf 87}, 047003 (2001).

\bibitem{Eis04}
H. Eisaki {\em et al.},
Phys. Rev. B {\bf 69}, 064512 (2004).

\bibitem{Li10}
L. Li {\em et al.}, 
Phys. Rev. B {\bf 81}, 054510 (2010).

\bibitem{Gom07} K.K. Gomes {\em et al.},
Nature {\bf 447}, 569 (2007).

\bibitem{Pas08}
A.N. Pasupathy {\em et al.},
Science {\bf 320}, 196 (2008).


\bibitem{McE05}
K. McElroy {\em et al.},
Science {\bf 309}, 1048 (2005).

\bibitem{Nun05}
T.S. Nunner, B.M. Andersen, A. Melikyan, and P.J. Hirschfeld, Phys.
Rev. Lett. {\bf 95}, 177003 (2005).

\bibitem{He06}
Y. He, T.S. Nunner, P.J. Hirschfeld, and H.-P. Cheng, Phys. Rev.
Lett. {\bf 96}, 197002 (2006).


\bibitem{Mas07}
M.M. Ma\'ska, Z. \'Sled\'z, K. Czajka, and M. Mierzejewski, Phys.
Rev. Lett. {\bf 99}, 147006 (2007).

\bibitem{Pet09} 
S. Petit and M.-B. Lepetit, EPL {\bf 87}, 67005 (2009).

\bibitem{Foy09}
K. Foyevtsova, R. Valent\'i, and P.J. Hirschfeld, Phys. Rev. B {\bf
79}, 144424 (2009).

\bibitem{Joh09}
S. Johnston, F. Vernay, and T.P. Devereaux, EPL {\bf 86}, 37007
(2009).

\bibitem{Oka10}
S. Okamoto and T.A. Maier, Phys. Rev. B {\bf 81}, 214525 (2010).

\bibitem{Mae04}
Chapter 2 in S. Maekawa {\em et al.},
{\em Physics of Transition Metal Oxides}
(Springer-Verlag, Berlin, 2004).

\bibitem{Ric06} 
P. Richard {\em et al.}, 
Phys. Rev. B {\bf 74}, 094512 (2006).

\bibitem{Boe84}
D.K.G. de Boer, C. Haas, and G.A. Sawatzky,
Phys. Rev. B {\bf 29}, 4401 (1984).

\bibitem{Bri95}
J. van den Brink {\em et al.},
Phys. Rev. Lett. {\bf 75}, 4658 (1995).


\bibitem{Mor08}
M. Mori, G. Khaliullin, T. Tohyama, and S. Maekawa, Phys. Rev. Lett.
{\bf 101}, 247003 (2008).

\bibitem{And78}
O.K. Andersen, W. Klose, and H. Nohl,
Phys. Rev. B {\bf 17}, 1209 (1978).

\bibitem{McM88} 
A.K. McMahan, R.M. Martin, and S. Satpathy, 
Phys. Rev. B {\bf 38}, 6650 (1988).



\bibitem{Sha93}
R.D. Shannon, J. Appl. Phys. {\bf 73}, 348 (1993).  

\bibitem{noteE} The $3s-2p$ level separation $E_s-E_p$ for a negatively 
charged O$^{2-}$ ion is expected to be less than its value $\simeq 10$~eV 
for a neutral oxygen atom.  


\bibitem{notet} This $d$-$d$ virtual hopping should not be confused 
with the Zhang-Rice singlet hopping $t$ in the $t-J$ model. 

\bibitem{note_tpd} The modulations of $t_{pd}$ and $\Delta_{pd}$ induced by 
structural distortions lead to only small changes in $J$ \cite{Joh09}. 

\bibitem{noteJ} A dopant also increases $J$ on the other 
closely located bonds, e.g., on 0--2, 0--9,... links (see inset in Fig.~4), 
but the effect is smaller and ignored here.  

\bibitem{Bra10} L. Braicovich {\em et al.}, 
Phys. Rev. Lett. {\bf 104}, 077002 (2010).

\end{thebibliography}
\end{document}